\documentclass{llncs}
\usepackage{amssymb,mathrsfs,amsmath,rotating}

\renewcommand{\Re}{{\mathbb{R}}}
\newcommand{\C}{{\mathbb{C}}}

\newcommand{\Z}{{\mathbb{Z}}}
\newcommand{\N}{{\mathbb{N}}}

\def\path{\begin{turn}{-95}$\scriptscriptstyle
\rightsquigarrow$\end{turn}}

\def\CPRp{{\rm \#^{\rm path}P_{\kern-2pt\Re}}}
\def\CPRo{{\rm \#^{\sf cc}P_{\kern-2pt\Re}}}
\def\CPRd{{\rm \#^{\rm d}P_{\kern-2pt\Re}}} 
\def\CPRi{{\rm \#^{\bullet}P_{\kern-2pt\Re}}}
\def\CPao{{\rm \#^{\sf cc}P_{\kern-2pt\rm add}}}
\def\CPad{{\rm D\#P_{\kern-2pt\rm add}}}
\def\CPai{{\rm \#P_{\kern-3pt\rm add}}}
\def\FPAR{{\rm FPAR}_{\kern-0.4pt\Re}}
\def\FPR{{\rm FP}_{\kern-1pt\Re}}
\def\PHa{{\rm PH}_{\kern0pt\rm add}}
\def\EXPadd{{\rm EXP_{\kern-2.5pt\rm add}}}

\def\P{{\sf P}}
\def\NP{{\sf NP}}

\def\NC{{\sf NC}}

\def\PSPACE{{\sf PSPACE}}
\def\LOGSPACE{{\sf LOGSPACE}}
\def\FLOGSPACE{{\sf FLOGSPACE}}

\def\EXP{{\sf EXP}}

\def\EXPadd{{\rm EXP}_{\!{\rm add}}}

\def\Po{{\rm P\kern-1.5pt_{\hbox{\eightrm add}}^{\,\leq}}}
\def\NPo{{\rm NP\kern-1.5pt_{\hbox{\eightrm add}}^{\,\leq}}}
\def\Peq{{\rm P\kern-1.5pt_{\hbox{\eightrm add}}^{\,=}}}
\def\NPeq{{\rm NP\kern-1.5pt_{\hbox{\eightrm add}}^{\,=}}}

\def\PAR{{\rm PAR}_{\Re}}

\def\FPAR{{\sf FPAR}}

\def\PAR{{\sf PAR}}

\def\FPR{{\rm FP}_{\kern-1pt\Re}}
\def\NPC{{\rm NP}_{\kern-2pt\C}}

\def\CNPad{{\rm \#^{\rm d}NP}_{\kern-2pt\rm add}}

\def\CDIMa{{\rm CDIM}_{\kern0pt\rm add}}
\def\CINFa{{\rm CINF}_{\kern-2pt\rm add}}

\def\CFEASadd{{\rm CFEAS}_{\kern-3pt\rm add}}
\def\FPadd{{\rm FP}_{\kern-3pt\rm add}}
\def\DNPadd{{\rm DNP}_{\kern-3pt\rm add}}

\def\FPARadd{{\rm FPAR}_{\kern0pt\rm add}}

\def\FEASR{{\rm FEAS}_{\kern-1pt\Re}}
\def\FEASC{{\rm FEAS}_{\kern-1pt\C}}

\def\FFEASR{{\rm 4FEAS}_{\kern-1pt\Re}}
\def\2SAS{{\rm 2SAS}_{\kern-1pt\Re}}

\def\CSATR{{\rm CSAT}_{\kern-1pt\Re}}
\def\CSATa{{\rm CSAT}_{\kern0pt\rm add}}
\def\EULERadd{{\rm EULER}_{\kern0pt\rm add}}
\def\BETTIadd{{\rm BETTI}_{\kern0pt\rm add}}
\def\EULER01{{\rm EULER}_{\kern0pt\rm add}^0}
\def\BETTI01{{\rm BETTI}_{\kern0pt\rm add}^0}
\def\STCa{{\rm STC}_{\kern0pt\rm add}}

\def\dpex{{\raisebox{1.1ex}{\mbox{\begin{turn}{180}%
${\boldsymbol{\epsilon}}$\end{turn}}}}_{_{\kern-1pt\sf D}}}

\begin{document}

\title{A Measure of Space for Computing over the Reals}

\titlerunning{A Measure of Space for Computing over the Reals}
%
\author{Paulin Jacob\'e de Naurois\thanks{Partially supported by the ANR project NO CoST: New tools for complexity - semantics and types}}

\authorrunning{P.~de Naurois}
\tocauthor{Paulin Jacob\'e de Naurois (LIPN - Universit\'e Paris XIII)}

\institute{LIPN - Universit\'e Paris XIII\\
99, avenue Jean-Baptiste Cl\'ement \\
93430 Villetaneuse - FRANCE\\
email: \email{denaurois@lipn.univ-paris13.fr}}

\maketitle              

\begin{abstract}
We propose a new complexity measure of space for the BSS model of computation. We define $\LOGSPACE_W$ and $\PSPACE_W$ complexity classes over the reals. We prove that $\LOGSPACE_W$ is included in $\NC^2_\Re \cap \P_W$, i.e. is small enough for being relevant. We prove that the Real Circuit Decision Problem is $\P_\Re$-complete under $\LOGSPACE_W$ reductions, i.e. that $\LOGSPACE_W$ is large enough for containing natural algorithms. We also prove that $\PSPACE_W$ is included in $\PAR_\Re$. 
\end{abstract}

\section*{Introduction}

The real number model of computation, introduced in 1989 by Blum, Shub and Smale in their seminal paper~\cite{BSS:89}, has proved very successful in providing a sound framework for studying the complexity of decision problems dealing with real numbers. A large number of complexity classes have been introduced, and many natural problems have been proved to be complete for these classes. A nice feature of this model is that it extends many concepts of the classical complexity theory to the broader setting of real computation; in particular a question $\P_\Re \neq \NP_\Re$ has arisen, which seems at least as difficult to prove as the classical one, and several $\NP_\Re$-complete natural problems have been exhibited.

It has been soon pretty obvious, however, that all features of the classical complexity theory could not be brought to this setting. In particular, the only complexity measures considered so far were dealing with {\em time}, and not, say, {\em space}: in 1989, Michaux proved in~\cite{M:89} that, under a straightforward notion of space, everything is computable in constant space. Therefore, no notion of logarithmic or polynomial space complexity exists so far over the reals. A way to deal with this situation has been to define parallel complexity classes in terms of algebraic circuits, such that the $\NC^i_\Re$ and the $\PAR_\Re$ classes. 

This model of computation has also long been criticized for being unrealistic: the assumption that one could multiply two arbitrary real numbers in constant time was the usual target, that Koiran faced in~\cite{K:97} by defining a notion of {\em weak} cost that increases the cost for repeatedly multiplying or adding numbers. 

Inspired by his approach, we propose here a new measure of space for the real number model, denoted as {\em weak} space, such that a repeated sequence of multiplications or additions on a number increases its size. Our notion allows us to define a logarithmic space complexity class, that falls within $\NC^2_\Re$. We also prove that this class is large enough for containing natural algorithms: in particular, we prove a $\P_\Re$-completeness result under $\LOGSPACE_W$ reductions.

The paper is organized as follows: in Section~\ref{sec:BSS}, we recall concepts and notations from the BSS model of computation. We define machines, circuits, and some major complexity classes. In Section~\ref{sec:Michaux}, we briefly recall Michaux's result, and sketch a proof. In Section~\ref{sec:Koiran}, we briefly introduce Koiran's notion of weak cost, and state some of the major results related to this notion. Then, we introduce our notion of {\em weak size} in Section~\ref{sec:size}, and state our results.

\section{A Short Introduction on the BSS Model}~\label{sec:BSS}

In this section, we list the notations used in the paper, and recall some basic notions and results on the BSS model. A comprehensive reference for these notions is~\cite{BCSS:98}.

\subsection{Notations}

For an integer $c \in \Z$ we define its height as $\lceil \log(|c|+1) \rceil$. The height of an integer is the number of digits of its binary encoding. We also define $\Re^* = \bigcup _{n\in\N} \Re^n$.

\subsection{Real Machines}

We consider BSS machines over $\Re$ as they
are defined in~\cite{BSS:89,BCSS:98}. Roughly speaking, such a machine takes
an input from $\Re^*$ , performs a number of arithmetic operations
and comparisons following a finite list of instructions,
and halts returning an element in  $\Re^*$ (or loops forever). Such a machine can be seen as a Turing machine over $\Re$. It essentially consists in a finite directed graph, whose nodes are \emph{instructions}, together with an input tape, an output tape, and a bi-infinite work tape, equipped with scanning heads. The instructions can be of the following types: \emph{Start}, \emph{Input} (reads an input value), \emph{Output} (writes an output value), \emph{Computation} (performs one arithmetical operation on two elements on the work tape), \emph{Constant} (writes a constant parameter $A_i\in\Re$), \emph{Branch} (compares two elements, and branches accordingly), \emph{Shift}, \emph{Copy} and \emph{Halt}.

For a given machine $M$, the function $\varphi_M$ associating its 
output to a given input $x\in\Re^*$ is called the \emph{input-output function}.  
We say that a function $f:\Re^*\to\Re^*$ is 
{\em computable} when there is a machine $M$ such that 
$f=\varphi_M$.

Also, a set $L\subseteq\Re^*$, or a \emph{language} is \emph{decided} by a machine 
$M$ if its characteristic function $\chi_L:\Re^*\to\{0,1\}$ 
coincides with $\varphi_M$.

This model of computation allows one to define complexity classes. In particular, $\P_\Re$ is the set of subsets of $\Re^*$ that are decided by a real machine that works in deterministic polynomial time. Similarly, $\NP_\Re$ is the set of subsets of $\Re^*$ that are decided by a real machine that works in nondeterministic polynomial time i.e, $x\in\Re^*$ is accepted if and only if there exists $y\in\Re^*$ of polynomial size such that the machine accepts $(x,y)$.

\subsection{Configurations}

\begin{definition} {\rm Configurations}
\begin{itemize}
\item A {\em configuration} of a machine $M$ is given by
an instruction $q$ of $M$ along with the position of 
the heads of the machine and three words $w_{input} \in \Re^*$, $w_{work}\in\Re_*$, $w_{output} \in \Re ^*$ that 
give the contents of the input tape, of the work tape and of the output tape.
\item A {\em transition} of a machine $M$ is a couple $(c_i, c_j)$ of configurations such that, whenever $M$ is in configuration $c_i$, $M$ reaches $c_j$ in one computation step.
\end{itemize}
\end{definition}

\begin{definition}\label{def:graph} {\rm Configuration Graph}\\
For a given machine $M$ and a given set ${\cal C}$ of configurations of $M$, we define the \emph{configuration graph} of $M$ on ${\cal C}$ to be the directed graph with vertexes all elements in ${\cal C}$, and edges all transitions of $M$ between elements in ${\cal C}$.
\end{definition}

\subsection{Algebraic Circuits}

We introduce the notion of algebraic circuits, that allows to denote parallel computations and to define complexity classes below $\P_\Re$.

\begin{definition} {\rm Algebraic Circuit}\\
An \emph{algebraic circuit} ${\cal C}$ is a sequence of gates $(G_1,\ldots,G_m)$ of one of the following types:
\begin{enumerate}
\item {\sl Input gates}: $G_i=x_i$, takes the input $x_i$ from $\Re$,
\item {\sl Arithmetic gates}: perform the operation $*$ to the outputs of gates $G_j$ and $G_l$, $j,l < i$ and $*\in\{+,-,.,/\}$,
\item {\sl Constant gates}: $G_i=A_i$, $A_i \in \Re$,
\item {\sl Sign gates}: If $G_j \ge 0$ then $G_i=1$ else $G_i=0$, $j<i$.
\end{enumerate}
If a circuit has $n$ {\sl input} gates, we can suppose that they are the first ones, $G_1,\ldots,G_n$. If moreover the last node $G_m$ is a {\sl sign} node, we shall say that ${\cal C}$ is a \emph{decision circuit}.
\end{definition}

An algebraic circuit is a finite directed graph with no loops: its \emph{size} is the number of gates, and its \emph{depth} is the length of its longest path, starting from an input gate.

Algebraic circuits extend the classical notion of boolean circuit to the BSS setting, and allows one to define the $\NC_\Re$ hierarchy of complexity classes:
\begin{definition} {\rm $\NC_\Re$}\\
For all $i\in\N$, $\NC_\Re^i$ is the class of real decision problems decided by a $\P$-uniform family of circuits of polynomial size and of depth bounded by ${\cal O}(\log^i(n))$, and
$$\NC_\Re = \bigcup _{i\in\N} \NC^i_\Re.$$
\end{definition}

As remarked by Poizat in~\cite{P:95}, in the definition above the uniformity of the family can be considered relative to the classical Turing model. Hence, a $\P$-uniform family of algebraic decision circuit is such that there exists an finite enumeration of all the constant gates in the family. There exists then a $\P$ time Turing machine which, on input $n,k$, outputs a discrete description of the $k^{\rm th}$ gate of the $n^{\rm th}$ circuit of the family.

\begin{proposition}~\cite{C:92}
$$\NC_\Re \subsetneq \P_\Re.$$
\end{proposition}

\section{Michaux's Result}\label{sec:Michaux}

This section is devoted to a brief exposition of Michaux's Result~\cite{M:89}, which states that a straightforward measure of space fails in differentiating one algorithm from another. In this section, we will use the following notion of space as a complexity measure:
\begin{definition} {\rm Unit Space}\\
Let $M$ be a machine over $\Re$, and let $c$ be a configuration of $M$. We define $USize(c)$, the \emph{unit size} of $c$ to be number of non-empty cells on the work tape at configuration $c$.
Assume that on an input $(x_1,\ldots,x_n)$, the computation of $M$ ends within $t$ computation steps. The computation follows a path $c_0,\ldots,c_t$. We define the \emph{unit space} used by $M$ on input $(x_1,\ldots,x_n)$ to be
$$USpace(M,(x_1,\ldots,x_n))=\max_{0\le k \le t} USize(c_k).$$
Assume that the running time of $M$ is bounded by a function $t$. We define the \emph{unit space} used by $M$ on input size $n$ to be
$$USpace(M,n)=\max_{(x_1,\ldots,x_n) \in \Re^n} USpace(M,(x_1,\ldots,x_n)).$$
\end{definition}

This notion of \emph{unit space} is essentially the same as the classical notion of space for Turing machines. While  in the classical Turing model this notion gives rise to a whole hierarchy of complexity classes like $\LOGSPACE$, $\PSPACE$, interlaced with the time hierarchy, this is not the case in the real setting. In order to precise a bit how unit space behaves on the reals, let us begin with the following well known technical result.

\begin{lemma}\label{lem:rat_frac}
Let $M$ be a real machine with parameters $A_1,\ldots,A_m$, whose running time is bounded by a function $t$. Let $n \in \N$. On any input $x_1,\ldots,x_n \in \Re^n$, at any computation step $k \le t(n)$, any non-empty cell on the work tape, say $e_l$, contains the evaluation of a rational fraction $f_{l,k} \in \Z(X_1,\ldots,X_{n+m})$ on $(x_1,\ldots,x_n,A_1,\ldots,A_m)$.
\end{lemma}

\begin{proof}
Details arguments can be found in~\cite{M:89,P:95,K:97}. We only sketch a proof here. The key argument is that, at any computation step $k$, the content of any cell $e_l$ is obtained from the input values and the parameter values by a finite sequence of arithmetical operations. Therefore, the value in $e_l$ is  the evaluation of a rational fraction $f_{l,k} \in \Z(X_1,\ldots,X_{n+m})$ on $(x_1,\ldots,x_n,A_1,\ldots,A_m)$.
\end{proof}

\begin{proposition}~\cite{M:89}
Let $L\subseteq \Re^*$ by a real language decided by a machine $M$ in time bounded by a function $t$. There exists a constant $k\in\N$ and a machine $M'$ deciding $L$ in unit space $k$.
\end{proposition}
\begin{proof}
We only sketch the proof here. The interested reader can find more explanations in~\cite{M:89,P:95}.

Rational fractions with integer coefficients can be easily encoded in binary, therefore, by Lemma~\ref{lem:rat_frac}, any configuration of $M$ can also be encoded in binary. It suffices to realize that this binary encoding can be embedded into the digits of only two real numbers. Then, there exists a machine $M'$ simulating $M$ with only a constant number of real registers, among which two are needed for encoding the configurations of $M$.

\end{proof}

\section{The Weak BSS Model by Koiran}\label{sec:Koiran}

\subsection{Definitions}

\begin{definition} {\rm Weak Cost}\\
Let $M$ be a machine whose running time is bounded by a function $t$, and let $A_1,\ldots,A_m$ be its real parameters. On any input $x_1,\ldots,x_n$, the computation of $M$ consists in a sequence $c_0,\ldots,c_t, t\le t(n)$ of configurations. To a transition $c_k, c_{k+1}$ in this sequence we associate its \emph{weak cost} as follows:
\begin{itemize}
\item If the current instruction of $c_k$ is a computation node, let $e_l$ be the current cell on the work tape: the transition $c_k, c_{k+1}$ consists in the computation of a rational fraction $f_{l+1,k+1}=g_{l+1,k+1}/h_{l+1,k+1} \in \Z(x_1,\ldots,x_n,A_1,\ldots,A_m)$, which is placed on the cell $e_{l+1}$ in $c_{k+1}$. The \emph{weak cost} of the transition $c_k, c_{k+1}$ is defined to be the maximum of deg($g_{l+1,k+1}$), deg($h_{l+1,k+1}$), and the maximum height of the coefficients of $g_{l+1,k+1}$ and $h_{l+1,k+1}$.
\item Otherwise, the \emph{weak cost} of the transition $c_k, c_{k+1}$ is defined to be 1.
\end{itemize}
The \emph{weak running time} of $M$ on input $x_1,\ldots,x_n$ is the sum of the weak costs of the transitions in the sequence $c_0,\ldots,c_t$.

The \emph{weak running time of $M$} is the function that associates with every $n$ the maximum over all $x\in\Re^n$ of the running time of $M$ on $x$.
\end{definition}

\subsection{Some Results}

\begin{lemma}\label{lem:koiran}\cite{K:97}
A function is polynomial-time in the weak BSS model if and only if it is polynomial-time computable in the standard BSS model and the rational fractions $f_{l,k}$ have polynomial degree and coefficients of polynomial bit-size.
\end{lemma}

Let $\P_W$ (respectively $\NP_W$) be the set of real languages decided in deterministic (resp. nondeterministic) weak polynomial time, and $\EXP_W$ be the set of real languages decided in weak exponential time by a real machine.

\begin{proposition}
$$\NC^2_\Re \not \subset \P_W \subsetneq \P_\Re \subseteq \NP_W=\NP_\Re \subseteq \PAR_\Re \subseteq \EXP_W.$$
\end{proposition}
$\P_W \subsetneq \NP_W=\NP_\Re$ is from~\cite{CSS:94}, 
$\NP_\Re \subseteq \EXP_W$ from~\cite{K:97} (where it is shown that the inclusion is strict). The missing items can be found in~\cite{BCSS:98}.

\section{Weak Size and Space}\label{sec:size}

\subsection{Definitions}

Instead of considering a unit size for all values on the work tape, which allows one to decide every decidable language in constant space, we would like to have a notion of size for the values computed on the work tape. The weak size of a computed value is a reasonable upper bound for the size of a boolean description of the corresponding rational fraction with integer coefficients. The weak size of a configuration is then the sum of the weak sizes of all computed values on the work tape in this configuration.

Yet, we need to precise a bit more the idea. Our purpose is to have a ``nice'' measure of space, allowing one to define a reasonable logarithmic space class. A trivial rational fraction like $f_1(X_1,\ldots,X_n, A_1,\ldots, A_m) = X_1$ has clearly a boolean description of size $1$, while, for describing $f_n(X_1,\ldots,X_n, A_1,\ldots, A_m) = X_n$, one would need $\lceil \log(n+1) \rceil$ digits (for encoding the variable index). It seems rather unsatisfactory that a logarithmic space configuration may have a logarithmic number of occurrences of $f_1$, but only a constant ones of $f_n$. This feature can be corrected by allowing a permutation of the input variables, provided the permutation is simple enough, i.e can be described in logarithmic boolean space. In this paper, we have restricted ourselves to circular permutations, that can be described by an offset in $\{0, \ldots, n-1\}$.

\begin{definition}\label{def:poly_weak_size} {\rm Weak Size }\\
Assume $A_1,\ldots, A_m \in \Re^m$ are given real numbers, and let $g\in\Z[X_1,\ldots,X_{n+m}]$ be a real polynomial with integer coefficients. Define a real polynomial $g_{A_1,\ldots,A_m}=g[X_1,\ldots,X_n, A_1,\ldots, A_m]$, with free variables $X_1,\ldots,X_n$. Let $0\le {\sf O} <n$, ${\sf O} \in \N$ be a number, the \emph{offset}.
To $g$, $A_1,\ldots, A_m \in \Re^m$ and {\sf O}, we associate the following:
\begin{itemize}
\item $deg(g)$ is the degree of $g$. We will write ${\sf D}(g)$ for $\lceil \log(deg(g) +1) \rceil$.
\item $Var_{A_1,\ldots,A_m}(g) \subseteq \{X_1,\ldots,X_n\}$ is the set of input variables on which $g_{A_1,\ldots,A_m}$ effectively depends.
\item $R_{A_1,\ldots,A_m, {\sf O}}(g) = \max \{i+{\sf O} \mod n\}$ for $X_i \in Var_{A_1,\ldots,A_m}(g)$, is the \emph{range} of $g$. We will write ${\sf R}(g)$ for $\lceil \log(R_{A_1,\ldots,A_m,{\sf O}}(g)+1) \rceil$.
\item ${\sf N}(g) \in \N$ is the number of non-zero monomials of $g$.
\item $S(g) \in \Z$ is the maximal absolute value of the integer coefficients of $g$. We will write ${\sf S}(g)$ for $\lceil \log(2S(g)+1) \rceil$.
\item $Vc_{A_1,\ldots,A_m}(g)$ is the maximum, for every monomial of $g$, of the number of input variables on which it effectively depends. We will write ${\sf V}(g)$ for $Vc_{A_1,\ldots,A_m}(g)$.
\end{itemize}

The weak size $S_{A_1,\ldots,A_m,{\sf O}}(g)$ of $g$ is defined as follows:
\begin{eqnarray}\label{eq:weak_size}
S_{A_1,\ldots,A_m,{\sf O}}(g) &=& {\sf N}(g) \left ( {\sf S}(g) + {\sf V}(g) . {\sf R}(g) + {\sf V}(g) . {\sf D}(g) \right )
\end{eqnarray}

For a rational fraction $f= g/ h$ we take the weak size of $f$ to be the maximum of the weak sizes of $g$ and $h$.
\end{definition}

It is clear that the weak size of $g$ bounds the size of a boolean encoding of $g$, where $g$ is presented as a sum of monomials modulo a circular permutation of the variable indexes. We do not take into account succinct boolean descriptions of factorized polynomials to ensure the tractability of our measure.

This measure of size for an element on the work tape naturally yields a notion of weak space for the given work tape, as follows:
\begin{definition}\label{def:conf_weak_size} {\rm Weak Space}\\
Let $M$ be a machine with real parameters $A_1,\ldots,A_m$. Let $c_k$ be a configuration of $M$, with the corresponding input $x_1,\ldots,x_n\in\Re^n$. We define:
\begin{itemize}
\item $e_{i},\ldots,e_{j}$ to be the non-empty part of the work tape in the configuration $c_k$.
\item For any non-empty cell $e_l$ in $c_k$, we denote by $f_{l,k}\in\Z(x_1,\ldots,x_n,A_1,\ldots,A_m)$ the rational fraction it contains.
\end{itemize}
The weak size of the work tape at the configuration $c_k$ is then:
$$Size_w(c_k) = \min_{0\le{\sf O}< n}\sum_{l=i} ^j S_{A_1,\ldots,A_m,{\sf O}}(f_{l,k})$$

Assume that the running time of $M$ is bounded by a function $t$. For a given input $x_1,\ldots,x_n$, the computation of $M$ consists in a sequence $c_0,\ldots,c_t, t\le t(n)$ of configurations. 

The \emph{weak running space} of $M$ on input $x_1,\ldots,x_n$ is the maximum for all configurations $c_0,\ldots,c_t$ of their weak size.

The \emph{weak running space} of $M$ is the function that associates with every $n$ the maximum over all $x\in\Re^n$ of the running space of $M$ on $x$.
\end{definition}

\newpage
\begin{definition} {\rm Complexity Classes}
\begin{itemize}
\item A language $L\subseteq\Re^*$ is in $\LOGSPACE_W$ if and only if there exist a machine $M$ and a constant $k\in\N$ such that, for all $n\in\N$, on input $x\in\Re^n$, $M$ decides whether $x\in L$ in weak space less than $k\log(n)$.
\item A language $L\subseteq\Re^*$ is in $\PSPACE_W$ if and only if there exist a machine $M$ and two constants $k,d\in\N$ such that, for all $n\in\N$, on input $x\in\Re^n$, $M$ decides whether $x\in L$ in weak space less than $kn^d$.
\item A function $f:\Re^* \rightarrow \Re^*$ is in $\FLOGSPACE_W$ if and only if there exist a machine $M$ and two constant $k,m\in\N$ such that, for all $n\in\N$, on input $x\in\Re^n$, and computation  $c_0,\ldots,c_t$ of $M$ on $x$:
\begin{enumerate}
\item $M$ computes $f(x)$ in weak space less than $k\log(n)$.
\item for every configuration $c_i$ with current node an {\sl output} node and current cell $e_l$, the weak size of the content of $e_l$ is less than $m$.
\end{enumerate}
\end{itemize}
\end{definition}

In the definition of $\FLOGSPACE_W$,  the output consists in a sequence of real values of constant weak size in the input. This ensures that one can compose $\FLOGSPACE_W$ algorithms, and that the result of the composition remains an $\FLOGSPACE_W$ algorithm. This is necessary for defining notions like logarithmic space reductions and for obtaining completeness results.

\subsection{What Michaux's Result Becomes}

\begin{lemma}\label{lem:size}
There exists $L\subseteq \Re^*$ such that:
\begin{itemize}
\item $L \in \P_W$
\item for all $k\in\N$, $L$ is not decidable in weak space less than $k$.
\end{itemize}
\end{lemma}

\begin{proof}
Let $p(X_1,\ldots,X_n) = X_1+\ldots +X_n$, and consider the set $L$ of points $(x_1,\ldots,x_n)\in\Re^n$, such that $p(x_1,\ldots,x_n)$ equals 0.
Assume $L$ is decided by a machine $M$. It is well known that the set of inputs accepted by a BSS machine is semi-algebraic, therefore, $L$ can be described as a finite union of sets given by systems of polynomials inequalities of the form
$$\bigwedge _{i=0} ^s F_i(X_1,\ldots,X_n)=0 \wedge \bigwedge _{j=0} ^t G_j(X_1,\ldots,X_n)>0,$$
where the values $F_i(x_1,\ldots,x_n)$ and $G_j(x_1,\ldots,x_n)$ are effectively computed by $M$.
Since $L$ has dimension $n-1$, at least one of these sets must have dimension $n-1$. Since the set described by the $G_j's$ is open, it must be nonempty, and then it defines an open subset of $\Re^n$.
All the polynomials $F_i$ vanish on that nonempty open subset of $L$. Since this open subset of $L$ is clearly infinite, and $p$ is an irreducible polynomial, all the polynomials $F_i$ must vanish on the whole set $L$. It is then a well known result~(\cite{BCSS:98}, Proposition 2 p.362) that the polynomials $F_i$ are multiples of $p$. Also, at least one of these $F_i$ is a non-trivial multiple of $p$.

It is clear that $p(x_1,\ldots,x_n)$ has weak size at least $n\log(n)$, and so does this non-trivial multiple of $p$. Therefore, $M$ decides $L$ in weak space at least $n\log(n)$.

\end{proof}

\subsection{Structural Complexity Results}

\begin{theorem}\label{prop:log_in_Pw}
\begin{eqnarray*}
\LOGSPACE_W &\subseteq& \P_W \cap \NC^2_\Re,\\
\PSPACE_W &\subseteq&\PAR_\Re.
\end{eqnarray*}
\end{theorem}

\begin{proof}

In a first step, we prove $\LOGSPACE_W \subseteq \P_\Re$. The key argument is an upper bound for the number of configurations of weak size at most $k\log(n)$. Consider a machine $M$, with $t$ nodes. For a fixed input size $n$, and an offset ${\sf O}$, simple counting arguments show that the number of rational fractions of weak size at most $B$ for some $B \in \N$ is bounded by $\alpha^B$, for some $\alpha\in\N$. It follows that the number of possible work tape contents of weak size $B$, for the same fixed offset, is bounded by $(2\alpha^2)^B$. Taking into account all possible values for the offset, the scanning head positions and the current node of the machine, the number of configurations of weak size at most $B$ is then bounded by $tn^2B(2\alpha^2)^B$. When $B=k\log(n)$, this bound is polynomial.

$\LOGSPACE_W \subseteq \P_W$ follows then by Lemma~\ref{lem:koiran}, since all rational fractions of logarithmic weak size have clearly polynomial degrees and coefficient heights.

$\LOGSPACE_W \subseteq \NC^2_\Re$ is then proven along the lines of~\cite{B:77}: given a $\LOGSPACE_W$ machine $M$, we exhibit a $\NC^1_\Re$ construction of its configuration graph. This construction involves some numeric computation, in order to check whether two given configurations are connected, and produces a boolean description of the configuration graph of $M$. Next, it suffices to decide whether the input and accepting configurations are connected in this graph: this is the classical reachability problem, which is decidable in the boolean class $\NC^2$.

$\PSPACE_W \subseteq\PAR_\Re$ is a corollary.
\end{proof}

\subsection{Completeness Results}

\begin{definition}~\cite{CT:92} {\rm Real Circuit Decision Problem (${\sf CDP}_\Re$)}\\
Input: $({\cal C}, \overline{x})$, where {\cal C} is an arithmetic circuit with $k$ input gates and $\overline{x} \in \Re^k$.\\
Question: Does ${\cal C}$ output 1 on input $\overline{x}$?
\end{definition}

It has been shown in~\cite{CT:92} that ${\sf CDP}_\Re$ is $\P_\Re$-complete under $\NC^2_\Re$-reductions.

\begin{theorem}\label{theo:complete}
${\sf CDP}_\Re$ is $\P_\Re$-complete under $\FLOGSPACE_W$-reductions.
\end{theorem}
 
\begin{proof}
The proof follows~\cite{CT:92}. The reduction happens to be in $\FLOGSPACE_W$.
\end{proof} 
We have stated this completeness results under $\FLOGSPACE_W$ reductions. By Theorem~\ref{prop:log_in_Pw}, it is clear that $\FLOGSPACE_W$ reductions are in $\P_W \cap \NC^2_\Re$. The problem considered has already been proven complete under first-order reductions~\cite{GrMe:96}, which also happen to be in $\P_W \cap \NC_\Re$. Yet, it remains unclear how the two types of reductions compare. 

\section{Concluding Remarks and Open Questions}

In the discrete model, space has proven to be a very relevant complexity measure. Many natural problems have been found in $\LOGSPACE$, and many others in $\NC^2$ whose membership in $\LOGSPACE$ is unclear. We believe that weak space may play the same role in the real setting. An argument in this direction is the following remark: consider a real algorithm that reads an input, normalizes it to $\{0,1\}$ with some step function, and applies a boolean $\LOGSPACE$ procedure. Real complexity analysis until now only allowed one to say that such a real algorithm belongs to $\P_W \cap \NC^2_\Re$: the algorithmic flavor behind it was lost. However, it is now clear that such an algorithm belongs to $\LOGSPACE_W$. An important task now is to exhibit some natural problems in $\LOGSPACE_W$. Others in $\NC^2_\Re$ or $\P_W$, not easily in $\LOGSPACE_W$, may also be of interest.

Structural results remain also to be found. In particular, it needs to be checked whether the following conjecture holds:
\begin{conjecture}
\begin{eqnarray*}
\NC^1_\Re & \not \subset& \LOGSPACE_W,\\
\LOGSPACE_W \subseteq \NC^1_\Re & \Rightarrow & \LOGSPACE \subseteq \NC^1.
\end{eqnarray*}
\end{conjecture}
Similar questions arise also for $\PSPACE_W$.
\section*{Acknowledgements}

We thank the anonymous referees for their helpful comments, and for pointing out references to the notion of first-order reductions.


\end{document}